\documentclass[pra,preprint,showpacs,preprintnumbers,amsmath,amssymb,floatfix]{revtex4}

\usepackage{graphicx}
\usepackage{bm}
\usepackage{url}

\begin{document}

\title{Significance of M2 and E3 transitions for $4p^54d^{N+1}$ and 
$4p^64d^{N-1}4f$ configuration metastable level lifetimes}

\author{R. Karpu\v{s}kien\.e}
\email{Rasa.Karpuskiene@tfai.vu.lt}

\author{P. Bogdanovich}

\author{R. Kisielius}
\affiliation{Institute of Theoretical Physics and Astronomy, Vilnius
University, \ A. Go\v{s}tauto 12,  Vilnius, LT-01108, Vilnius, Lithuania}

\begin{abstract}

Magnetic quadrupole and electric octupole transitions from the configurations
$4p^54d^{N+1}$ and $4p^64d^{N-1}4f$ were calculated along with magnetic dipole, 
electric dipole and electric quadrupole radiative transitions 
in quasirelativistic Hartree-Fock approximation. 
Their significance in determining the metastable level radiative lifetimes was 
investigated along several isoelectronic sequences for the ions from $Z=50$ to 
$Z=92$. Strontium-like ions, zirconium-like ions, molybdenum-like ions and 
rhodium-like ions were studied comprehensively. Remaining isoelectronic 
sequences with the ground configuration $4d^{N}$ ($N=1,3,5,7,8,10$) were also 
reviewed albeit in less detail. A systematic trends of determined total 
radiative lifetimes were studied. The importance of magnetic quadrupole and 
electric octupole transitions from metastable levels of ions from these 
isoelectronic sequences was investigated and discussed. Inclusion of such 
transitions of higher multipole order can change theoretical radiative lifetime
values for some levels more than ten times. These cases can not be established 
in advance, without `performing detailed calculations.
 
\end{abstract}


\maketitle

\linespread{1.6}

\section{Introduction}

Theoretical investigation of spectral parameters for multicharged tungsten ions 
with an open 4d shell \cite{pb12,pb13} has clearly demonstrated that the 
electric octupole (E3) and magnetic quadrupole (M2) transitions from some 
levels of the excited configurations $4p^54d^{N+1}$ and $4p^64d^{N-1}4f$ to the
ground configuration $4p^64d^N$ play very important role in determining their 
lifetimes. This is caused by the fact that these particular levels are 
metastable ones with high values of total angular momentum $J$. Hence the 
electric dipole (E1) transitions from these levels to the ground configuration 
levels are not allowed by selection rules for $J$. Furthermore, in case when the
magnetic dipole (M1) or electric quadrupole (E2) transitions from these levels 
are weak, their radiative lifetimes $\tau$ are strongly influenced by M2 and E3 
transitions to the ground configuration levels.

Transitions of higher multipole order, such as the M2 and E3 radiative
transitions, previously were considered in \cite{biemont2004,cff06,safr06,safr09}. 
Extensive investigation for the elements ranging from Na-like to Ar-like was 
presented in \cite{cff06} where the M2 transitions were computed for magnesium,
sulfur and argon isoelectronic sequences.
The calculation of Ar-like ions was extended to include E3 transitions. 
It is easy to explain since there are only few levels with quite 
high $J$. Various radiative transitions in Ni-like ions were thoroughly 
studied in \cite{safr06}, including magnetic octupole transitions. 
Although this paper had presented a huge number of transitions, the cases when 
the M2 and E3 transitions were significant had not been underlined. 

The higher multipole-order radiative transitions were also calculated for 
$4f^{13}nl$ levels in \cite{safr09}, but their influence on the 
calculated radiative lifetimes was  not revealed. In general, the mentioned 
works were not specifically dedicated to the investigation of higher 
multipole-order transitions and did not involve comprehensive study of these 
transitions. Furthermore, it should be noted, that some experimental studies 
\cite{lun07, lun08, bie07} have indicated possible contribution of the
E2, M1, E3 and M2 transitions to the metastable level radiative lifetimes 
$\tau$.

In present work we investigate the influence of M2 and E3 transitions on the 
calculated radiative lifetimes of metastable levels for a wide range of
ions, starting with $Z = 50$ and going up to $Z = 92$ in isoelectronic
sequences with the ground configuration $4d^N$. We present comprehensive study
for four isoelectronic sequences with $N=2,4,6,9$. The calculations for 
remaining open-$d$ shell sequences are performed also, but the results are 
discussed only in brief. 

As a measure of significance of the radiative E2 and M3 transitions, we 
introduce parameter 
\begin{equation}
R = \tau_{\mathrm{E2+M1}}/\tau_{\mathrm{TOT}},
\end{equation} 
where $\tau_{\mathrm{TOT}}$ is a total radiative lifetime of a level determined 
from all (E2, M1, E3, M2) transition probabilities (E1 transition is forbidden 
for these levels), $\tau_{\mathrm{E2+M1}}$ is a radiative lifetime determined 
from transitions occurring inside the excited configurations $4p^54d^{N+1}$ and 
$4p^64d^{N-1}4f$ complex. The ratio $R$ shows how much a theoretical radiative 
lifetime decreases, when M2 and E3 transitions are included to $\tau$ 
calculation. Here we must underline that all possible transitions with their
probability values $A \le 10^{-12}$ from the levels under considerations are
included while determining $\tau$ values.

The radiative transition calculations were performed in a quasirelativistic 
approximation \cite{pbor06,pbor07}. 
The correlation corrections were not included in our calculations, 
because the main purpose of current work was to determine what kind of 
transitions are imperative in determining radiative lifetimes of excited 
levels rather than to calculate very accurate and reliable parameters of
radiative transitions.
Therefore the ground configuration $4p^64d^N$ was investigated in a 
one-configuration approximation. Furthermore, only the interaction between 
two excited configurations $4p^54d^{N+1}$ and $4p^64d^{N-1}4f$, which is very 
strong in multicharged ions, was included for the odd-parity states. 

\section{Calculation method}

We use a quasirelativistic Hartree-Fock approximation (QRHF) in our 
{\sl ab initio} calculations of ion energy levels and radiative transition
parameters, such as transition wavelengths $\lambda$, line strengths $S$, 
oscillator strengths $f$, transition probabilities $A$. In this approach, the 
one-electron radial orbitals $P(nl|r)$ are obtained by solving the 
quasirelativistic equations having the following form:

\begin{align}
\label{2.1}
&\left\{ \frac{d^2}{dr^2}-\frac{l(l+1)}{r^2}- V(nl|r)- \varepsilon_{nl} 
\right\} P(nl|r) - 
\nonumber \\
&X(nl|r) + \frac{\alpha^2}{4} \left( \varepsilon_{nl}+V\left(nl|r\right) \right)^2 
P(nl|r) +
\nonumber \\
&\frac{\alpha^2}{4} \left( \varepsilon_{nl}+V(nl|r) \right) X(nl|r) + 
\nonumber \\
&\frac{\alpha^2}{4} \left( 1- \frac{\alpha^2}{4} \left( \varepsilon_{nl}+V(nl|r) 
\right) \right)^{-1}D(nl|r)P(nl|r) = 0.
\end{align}

The first two lines of this equation represents the traditional Hartree-Fock 
equations, where $X(nl|r)$ denotes the exchange part of the potential and 
$V(nl|r)$ represents the direct part of the potential including the interaction 
of an electron with nucleus $U(r)$ and with other electrons. We take into 
account the finite size of a nucleus within the nuclear potential $U(r)$ 
\cite{pbor02,pbor03}. This allows us to express the radial orbitals in powers 
of a radial variable in the nucleus region. Next two terms with the 
multiplier $(\epsilon_{nl} + V(nl|r))$ describe the relativistic correction 
of the mass-velocity dependence. The last term of equation represents the 
potential of the electron contact interaction with nucleus. 

In our approach we include the contact interaction with the nucleus not only 
for the $s$ electrons but also  some part of that interaction for the 
$p$ electrons \cite{pbor07}:

\begin{align}
\label{2.2}
D(nl|r) = \left( \delta(l,0)+\frac{1}{3}\delta(l,1) \right)
\frac{dU(r)}{dr} 
\nonumber \\
\left( \frac{d}{dr} - \frac{1}{r} 
\left( \alpha^2 Z^2 \delta(l,1) \left( - \frac{37}{30}
- \frac{5}{9n} + \frac{2}{3n^2} \right) + 1 \right) \right) .
\end{align}

A detailed discussion of the particular features of main equation is given in 
\cite{pbor06,pbor07}, whereas their solution techniques are described in 
\cite{pbor03,pbvjor05}. 
Concluding the description of the employed approximation, we want to emphasize
that our quasirelativistic Hartree-Fock method significantly differs
from widely used approach described in \cite{hfr}. The main differences arise 
from our adopted set of quasirelativistic Hartree-Fock equations (QRHF) 
featuring several distinctive properties which are described in more detail
elsewhere \cite {pbor06,pbor07,pbor03,pbvjor05}.

The methods to calculate the energy level spectra were discussed extensively 
in \cite{pbor08}. For the energy level spectra calculation, we include all 
two-electron interactions in the same way as it is done in conventional
Breit-Pauli approximation. This similarity makes it possible to apply widely 
used code {\sc mchf breit} \cite{cff91a} for angular integration of
Breit-Pauli Hamiltonian 	matrix elements. We adopt computer program 
{\sc mchf mltpol} \cite{cff91b} to determine the 	matrix elements of transition 
operators along with the code {\sc mchf lsjtr} \cite{cff91c}, which has been 
adopted for use with the quasirelativistic radial orbitals.

\section{Results}

One can define 10 isoelectronic sequences with the ${\mathrm 4d^{N}}$ shell. 
In present work we have investigated isoelectronic sequences with two ($N=2$, 
strontium-like ions), four ($N=4$, zirconium-like ions), six ($N=6$, 
molybdenum-like ions) and nine ($N=9$, rhodium-like ions) electrons in the 
$4d$ shell of the ground configuration very thoroughly.
Remaining isoelectronic sequences with open ${\mathrm 4d^{N}}$ shell were 
examined also. Since the properties (and the behavior of the significance
parameter $R$) are very similar, we do not present a detailed analysis of 
isoelectronic $4d^N$ ($N=1,3,5,7,8,10$) sequences in current work.
 
We have included ions with comparatively high $Z$ values in order to avoid 
strong interaction between $4d$ and $5s$ shell electrons. Such an interaction 
is more significant for elements in low-level ionization degrees. 
It is obvious that one needs to perform complex calculations with correlation 
corrections included by adopting extensive configuration bases in order to
determine reliable data for radiative transitions or radiative lifetimes.
We did not perform this kind of calculations in present study because our
main purpose was to examine how inclusion of higher-multipole order transitions 
changes values of radiative lifetimes $\tau_{\mathrm{TOT}}$. It was not an
objective of current work to determine high-accuracy $\tau$ values, therefore 
we did not include correlation corrections in our study. Moreover, we have to 
point out that the influence of correlation corrections along isoelectronic 
sequence becomes less significant with increase of $Z$. 

For all isoelectronic sequences studied in current work, the levels of excited 
configurations, which have radiative lifetimes affected by M2 and E3 transitions
to the ground configuration, retain the same relative positions in energy level 
spectra. Nevertheless, their identification, based on the maximum contribution
from a particular $LS$ term, can be different. Therefore, this identification 
in $LS$ coupling scheme is performed for the low-level ionization ion with 
$Z = 56$.

\subsection{Sr-like ions}
\label{sr}

\begin{figure}[!ht]
\includegraphics[scale=0.52]{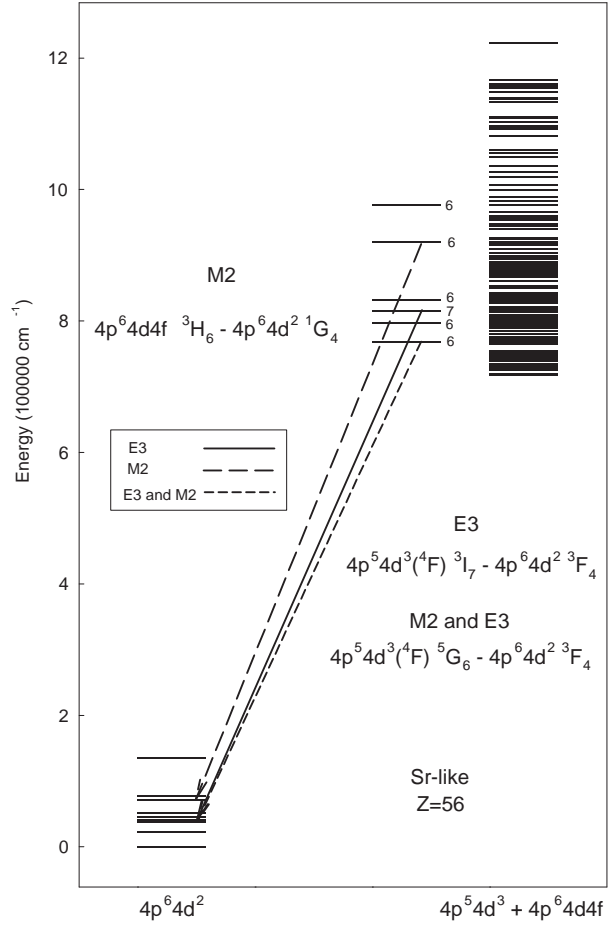}
\caption{
     \label{fig1} 
					Energy levels, E3 and M2 transitions for strontium-like barium ion
}
\end{figure}

Figure~\ref{fig1} demonstrates energy levels  of investigated configurations 
for the strontium-like ion  with $Z = 56$. The excited configuration levels 
with high total angular momentum $J$ values are presented in a separate column,
to the left from the column which shows all energy levels of two excited 
odd-parity configurations. Hence one can see the location and the quantity of 
these metastable levels in complete energy level spectra. The E1 transitions 
are not allowed from these levels. As a rule, the M1 and E2 transitions are 
allowed from these levels to the lower levels of excited configurations, 
and M2 and E3radiative transitions to the ground configuration are allowed too.
In general case, the M1 transition probability values are the largest ones, 
whereas the E3 transition probabilities are the weakest ones. Therefore, there 
exists assumption that the M1 transitions become the most important ones in 
radiative lifetimes $\tau_{\mathrm{TOT}}$ calculation, when E1 transitions are 
not allowed. In present work, we will investigate properties of such levels 
further and will demonstrate that this is not a correct estimate for all cases.

There are six energy levels with $J=6$ and one level with $J=7$ ($J$-values are 
given beside corresponding levels) in Fig~\ref{fig1}. The ground configuration
$4p^6 4d^2$ has only levels with $J=0,1,2,3,4$. The transitions, which are 
significant for the radiative lifetimes $\tau_{\mathrm{TOT}}$ and are discussed 
in Sect.~\ref{sr} also have been presented in Fig~\ref{fig1}. 

For the ions from the strontium isoelectronic sequence, we analyze the radiative 
lifetimes of three levels. The lifetimes of one of these levels, namely 
$4p^64d4f\,\,^3H_6$, are mostly affected by the M2 transition, although E3 
transitions are possible here too. The lifetimes of the second level with $J=7$
are affected by the E3 transition, because M2 transitions are not allowed from 
this level. The third level is more particular one, because its lifetimes are 
affected by both the M2 and E3 transitions. The transition 
$4p^54d^3\,(^4F)\,\,^5G_6 - 4p^64d^2\,\,^3F_4$, presented in Fig.~\ref{fig1}
by short-dashed line, demonstrates a particular case, when the excited level 
can decay not only by M1 transition, but by two additional different-type 
radiative transitions which have similar transition rate $A$ values.

Here and for other isoelectronic sequences, we present only those transitions, 
which are most significant for determined radiative lifetimes. Nevertheless, 
we must underline that all available transitions were included into calculation
of radiative lifetimes $\tau_{\mathrm{TOT}}$. All possible E2 and M1 transitions
were included in $\tau_{\mathrm{E2+M1}}$ calculation, too. However, one must 
keep in mind that the E2 transition probabilities $A$ usually are much weaker 
than those of M1 transitions. Therefore, the allowed and probable decay channel 
is attributed to the M1 transitions. In this and further sections, when we refer
to allowed (probable) transitions, we mean only the most probable transitions - 
two or more with the largest transition  probability values.

\begin{figure}[!ht]
\includegraphics[scale=0.52]{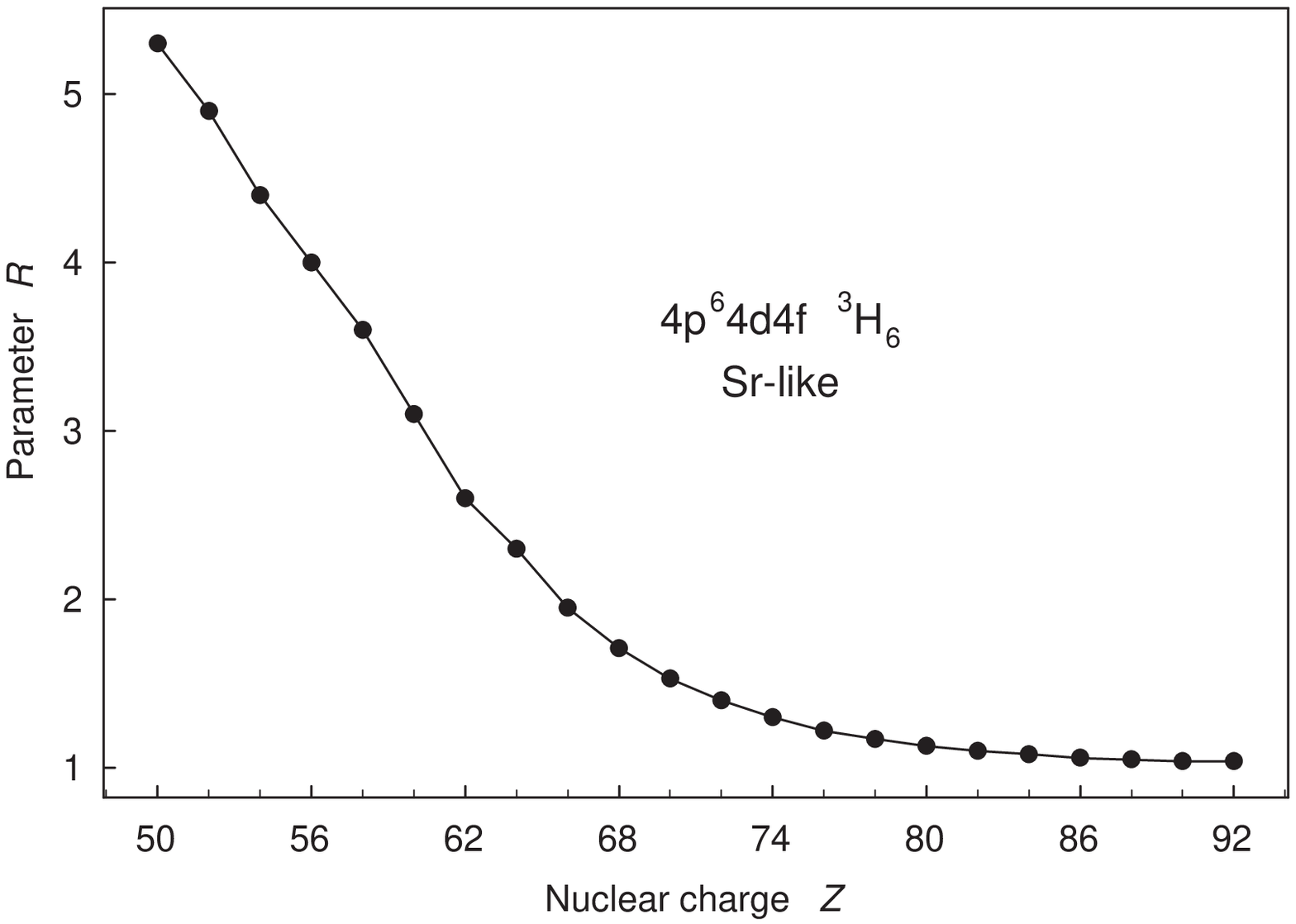}
\caption{
     \label{fig5} 
					Dependence of parameter $R$  on nuclear charge $Z$ for the level 
					$4p^64d4f\,\, ^3H_6$ of strontium-like ions
}
\end{figure} 

\begin{figure}[!ht]
\includegraphics[scale=0.52]{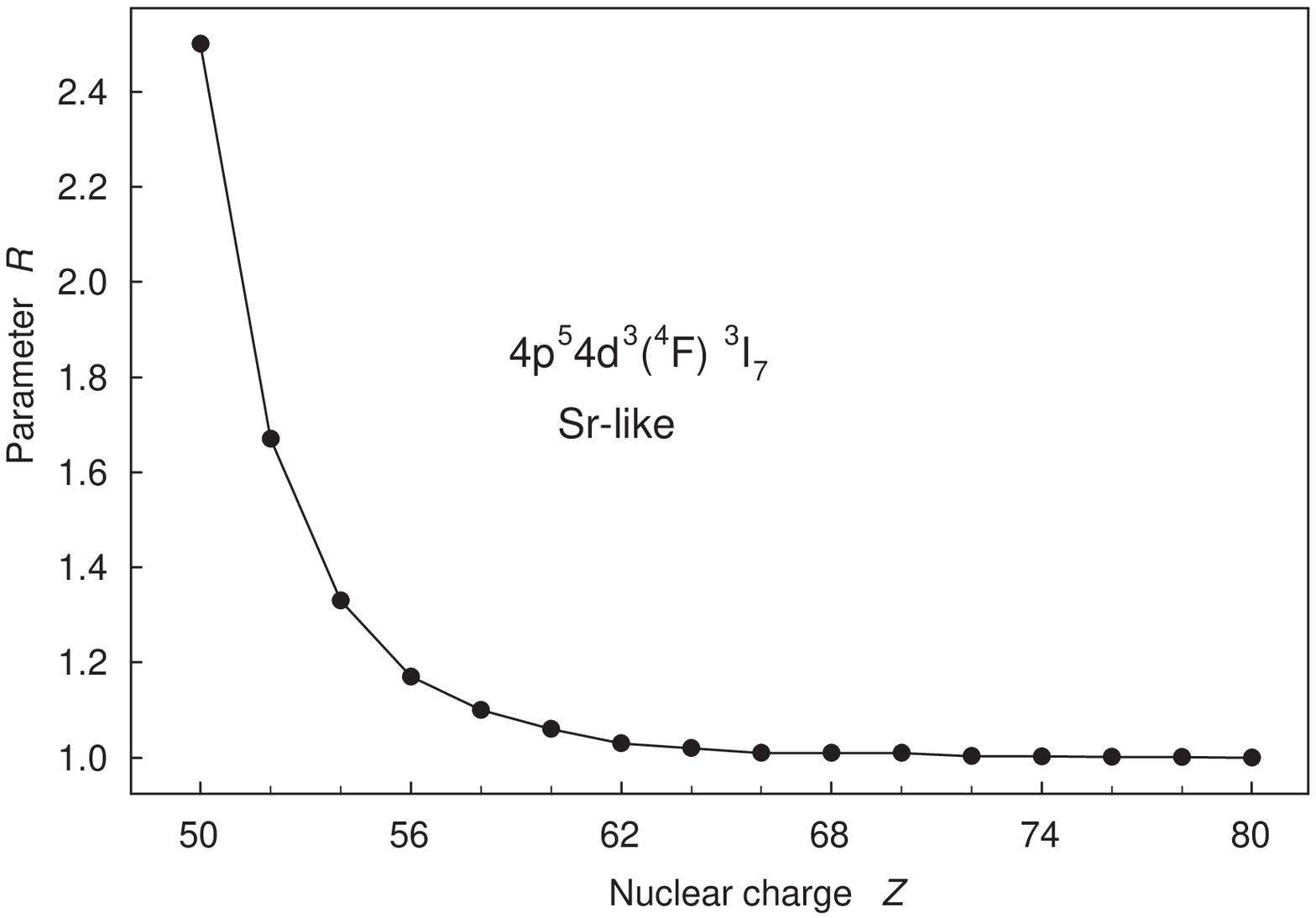}
\caption{
     \label{fig6} 
					Dependence of parameter $R$  on nuclear charge $Z$ for the level 
					$4p^54d^3\,( ^4F)\,\, ^3I_7$ of strontium-like ions
}
\end{figure} 

\begin{figure}[!ht]
\includegraphics[scale=0.52]{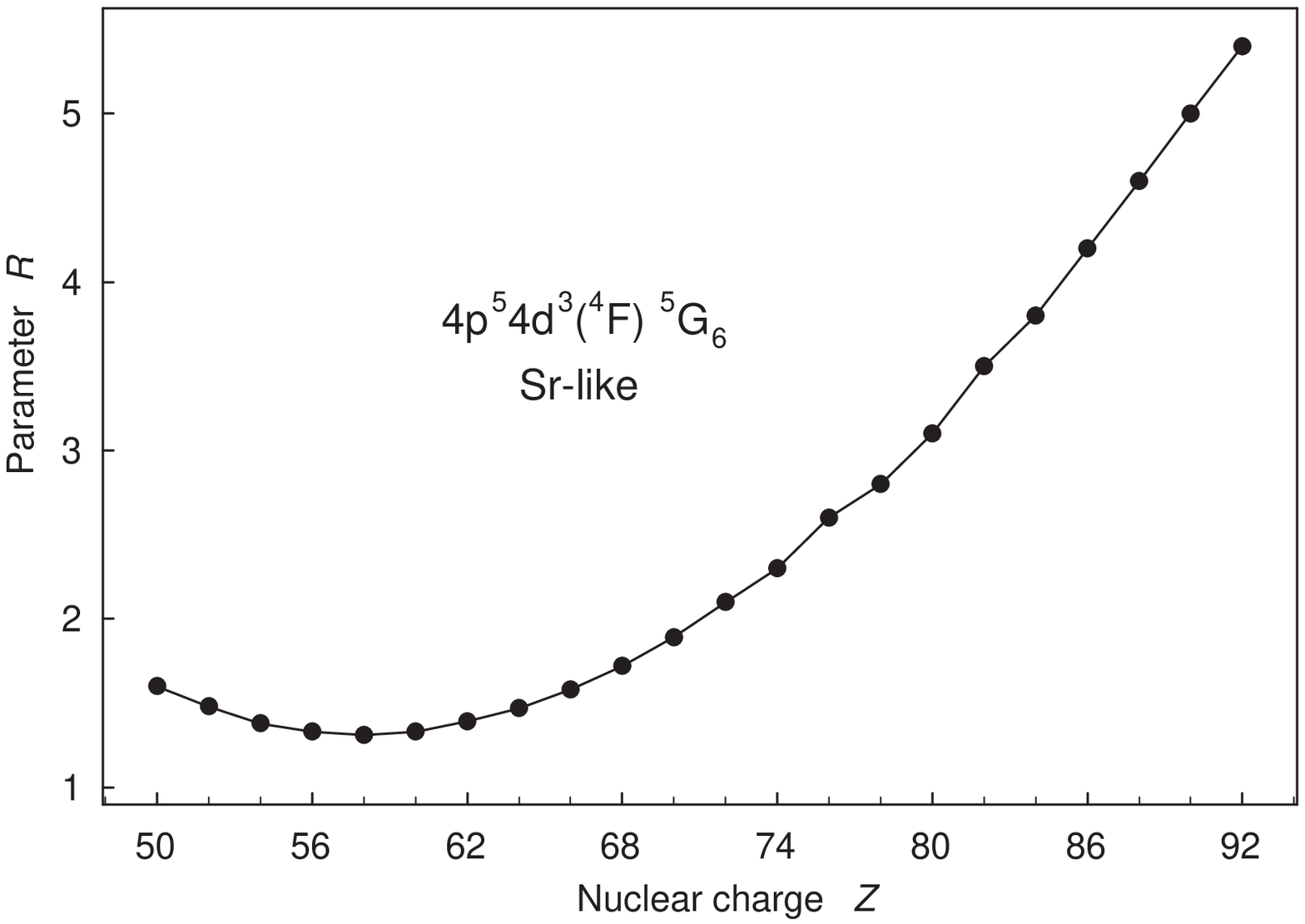}
\caption{
     \label{fig7} 
					Dependence of parameter $R$  on nuclear charge $Z$ for the level 
					$4p^54d^3\,( ^4F)\,\, ^5G_6$ of strontium-like ions
}
\end{figure} 

\begin{figure}[!ht]
\includegraphics[scale=0.52]{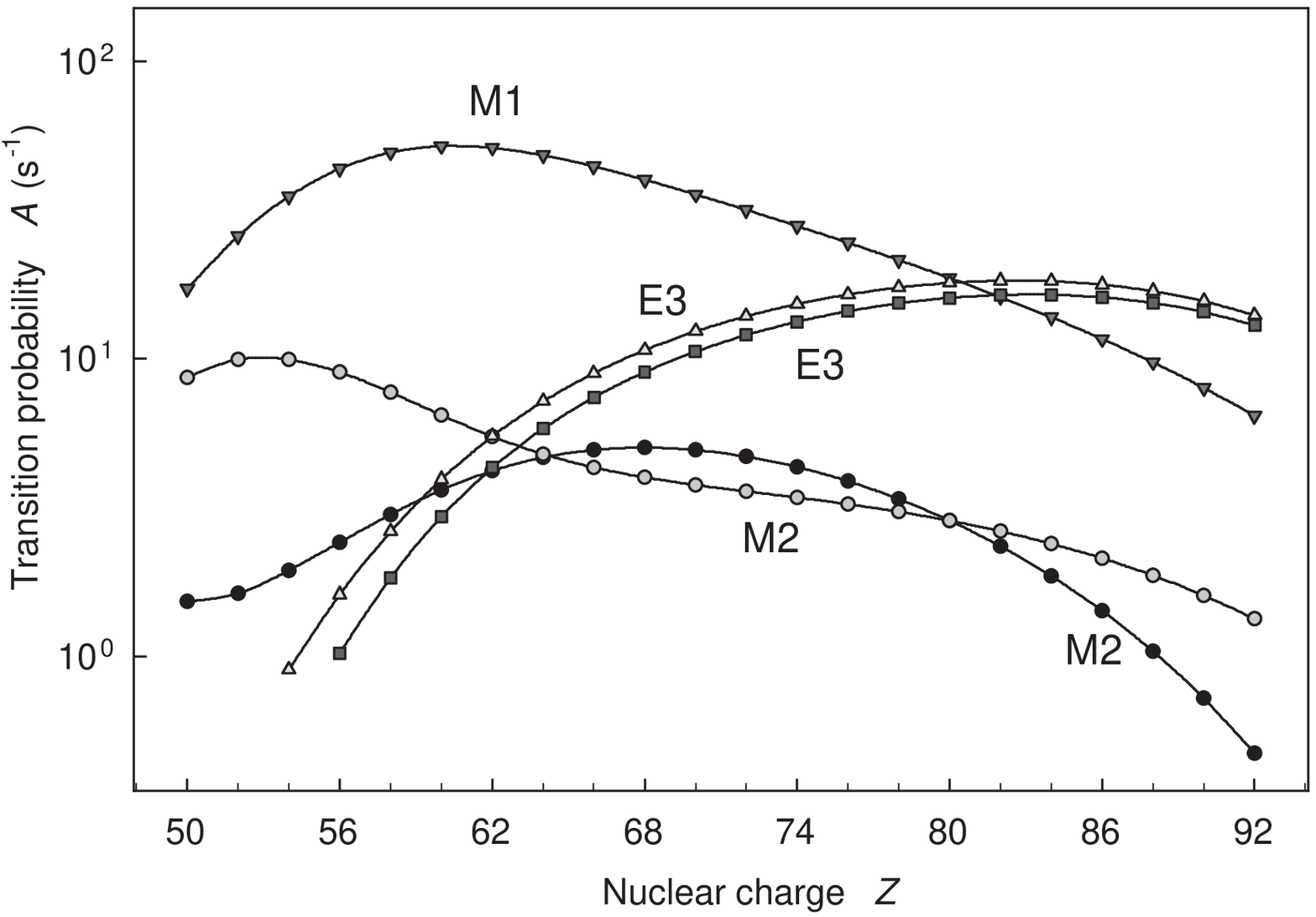}
\caption{
     \label{fig8} 
					Radiative transition probabilities originating from the 
					level $4p^54d^3\,( ^4F)\,\, ^5G_6$ along the strontium isoelectronic 
					sequence. M1 transitions are presented by grey triangles, two M2
					transitions are presented by grey and black circles, E3 transitions 
					are presented by white triangles and black squares.					
}
\end{figure}

Dependence of the parameter $R = \tau_{\mathrm{E2+M1}}/\tau_{\mathrm{TOT}}$ on 
a  nuclear charge $Z$ for the level $4p^64d4f\,\,^3H_6$ in the strontium 
isoelectronic sequence is presented in Fig.~\ref{fig5}, and for the level 
$4p^64d4f\,\,^3I_7$, it is given in Fig.~\ref{fig6}.  The parameter $R$ is 
significantly larger than 1 at the beginning of isoelectronic sequence, but it 
decreases fast as the nuclear charge $Z$ increases for these levels as well as 
for some  other levels not presented in plots. In the first case, for the level 
with $J = 6$, one or two M2 transitions are the most significant. For the next 
level with $J = 7$, this behavior is caused by E3 transition. The M2 transitions
are not allowed from this levels because the maximum value of total angular 
momentum $J = 4$ for the ground $4d^2$ shell. We have cut the parameter $R$ 
dependence in Fig.~\ref{fig6} at $Z=80$ because it is very close to 1 for 
higher $Z$, meaning that the E3 transitions are not significant any more.

Completely different dependence of the parameter $R$ on nuclear charge $Z$ 
is presented in Fig.~\ref{fig7} for the  level 
$4p^54d^3\,( ^4F)\,\, ^5G_6$ of the ions from the same isoelectronic 
sequence. Several decay channels: magnetic dipole transition M1 and two 
electric octupole transitions E3 as well as two magnetic quadrupole 
transitions M2, are important for this level. The dependences of the mentioned 
radiative transition probabilities on nuclear charge $Z$ are presented in 
Fig.~\ref{fig8}. Since the probability values change very significantly, 
a logarithmic scale is used for them here and in later plots. We do not provide
identification for the final levels of these transitions deliberately, because 
the identification of them as well of the upper level can change along 
sequence due to a strong mixing of $LS$ terms.

At the beginning of the sequence, where the main decay channel is M1 transition, 
the parameter $R$ decreases. For higher $Z$ values, the M1 transition 
probability as well as the M2 transition probabilities decrease whereas the E3 
transition probabilities increase very rapidly and, when $Z > 80$, these 
transition probabilities become most prominent and are most significant for the 
calculated radiative lifetimes $\tau_{\mathrm{TOT}}$. 

We want to add that there are other metastable levels with the parameter $R$ 
values and dependence on nuclear charge $Z$ similar to that presented in 
Fig.~\ref{fig5}.

\subsection{Zr-like ions} 
\label{zr}

\begin{figure}[!ht]
\includegraphics[scale=0.52]{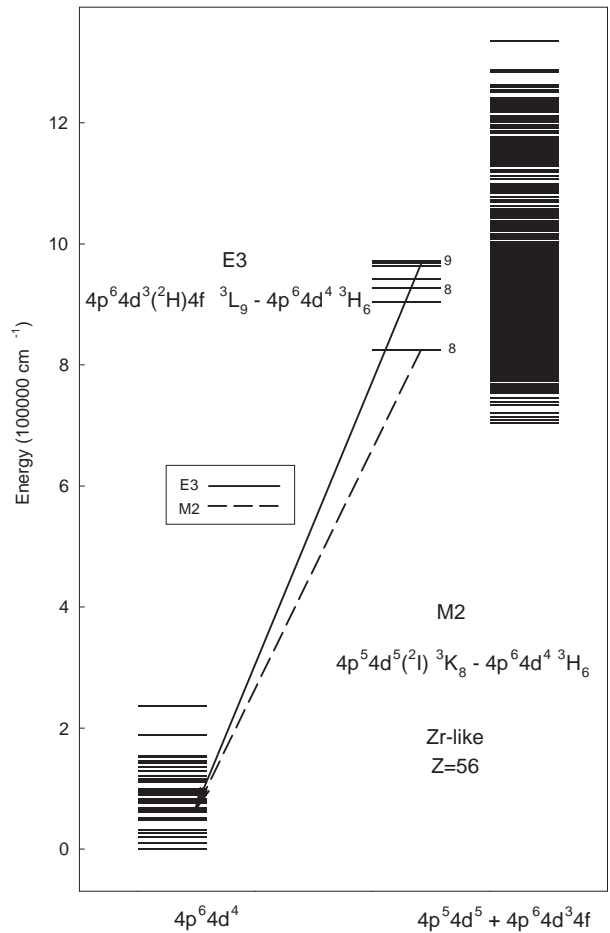}
\caption{
     \label{fig2} 
					Energy levels, E3 and M2 transitions for zirconium-like barium ion
}
\end{figure}

Figure~\ref{fig2} presents energy levels of investigated configurations for the
zirconium-like ion with $Z = 56$. The excited configuration levels with high 
total angular momentum $J$ values are presented in a separate column, 
to the left from the column which gives all energy levels of the odd-parity 
configurations like in Fig.~\ref{fig1}.  The E1 transitions are not allowed 
from these levels.  For the Zr-like ions, we have investigated six energy 
levels with $J = 8$ and one level with $J = 9$ originating from the complex of 
configurations $4p^54d^5 + 4p^64d^34f$. The ground configuration $4p^64d^4$ for
these ions can have the fine-structure levels with $J=0,1,2,3,4,5,6$. 
The transitions from metastable levels are presented in Fig.~\ref{fig2} and are 
discussed further in this section.

\begin{figure}[!ht]
\includegraphics[scale=0.52]{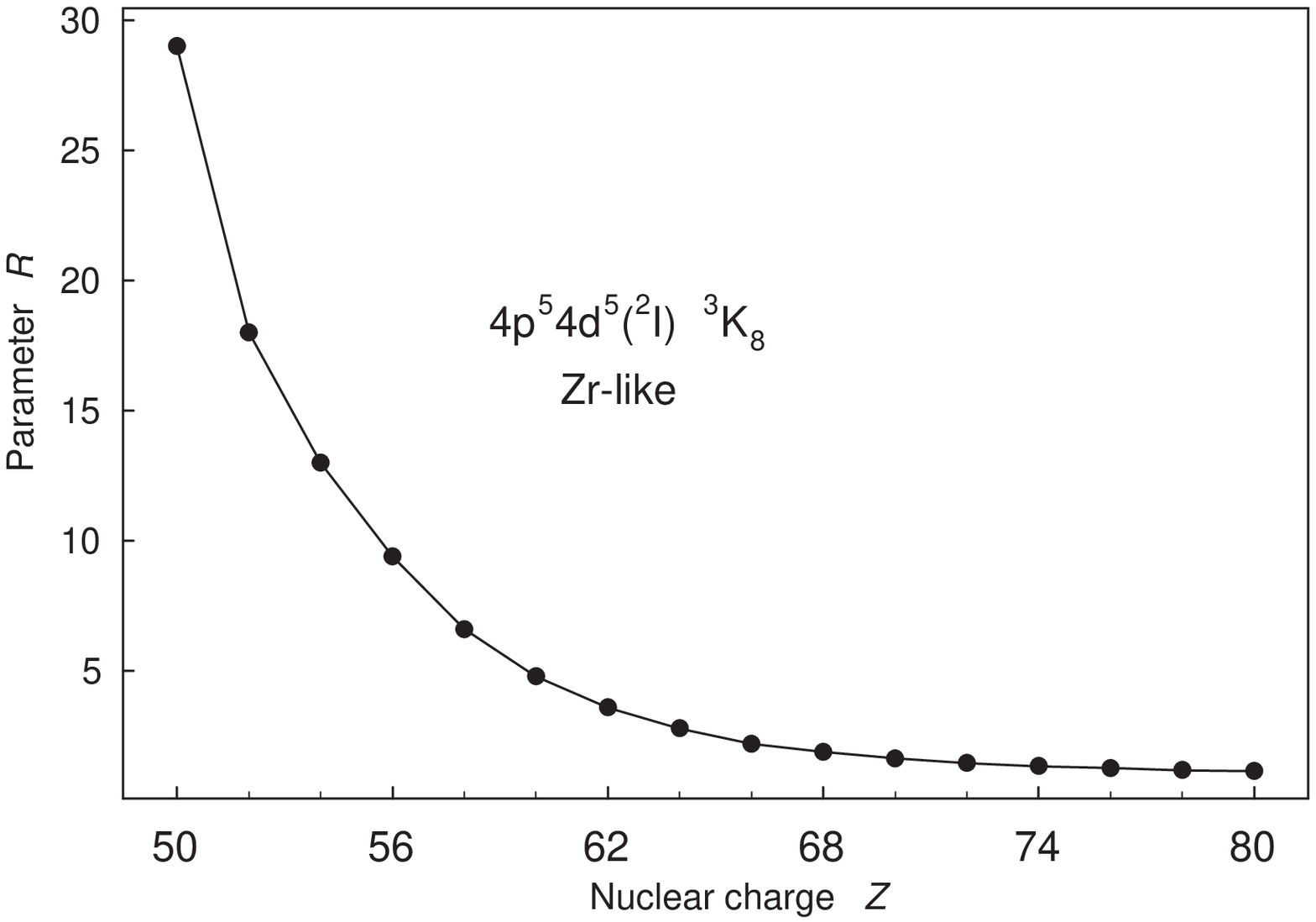}
\caption{
     \label{fig9} 
					Dependence of parameter $R$  on nuclear charge $Z$ for the level 
					$4p^54d^5\,( ^2I)\,\, ^3K_8$ of zirconium-like ions
}
\end{figure} 

\begin{figure}[!ht]
\includegraphics[scale=0.52]{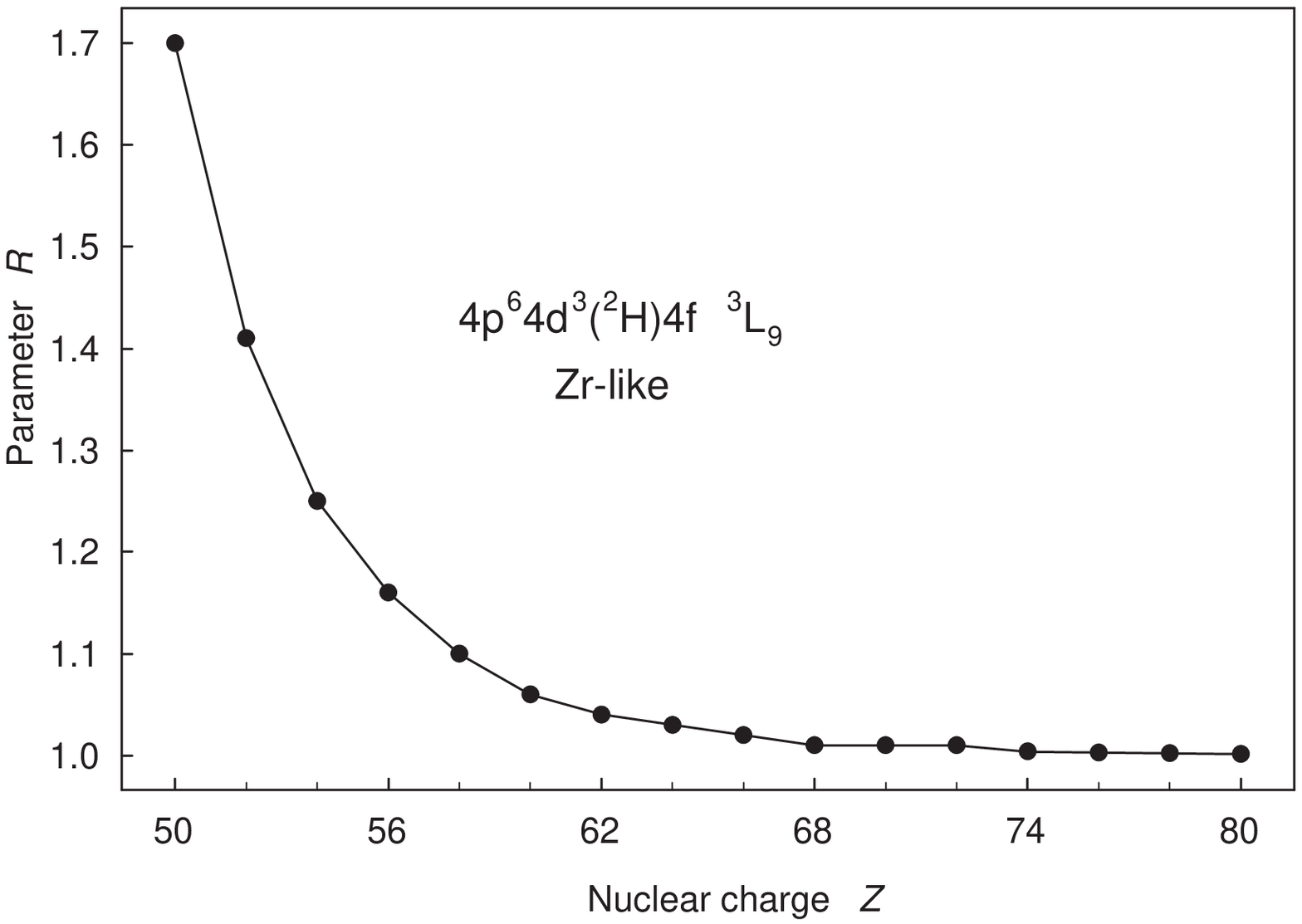}
\caption{
     \label{fig10} 
					Dependence of parameter $R$  on nuclear charge $Z$ for the level 
					$4p^54d^3\,( ^2H)4f\,\, ^3L_9$ of zirconium-like ions
}
\end{figure}

Figures~\ref{fig9} and \ref{fig10} show typical behavior of the parameter 
$R$ for the levels of ions from zirconium isoelectronic sequence. 
Figure~\ref{fig9} demonstrates the influence of the M2 transition on the 
determined radiative lifetime $\tau_{\mathrm{TOT}}$ for the level 
$ 4p^54d^5\,( ^2I)\,\, ^3K_8$. That influence is quite formidable at 
the beginning of the isoelectronic sequence. For other levels with $J = 8$, 
the parameter $R$ is similar in its value and behavior as in the case
of the $ 4p^54d^5\,( ^2I)\,\, ^3K_8$ level, 
even if these levels belong to another excited configuration, namely  
$4p^64d^34f$. We have dropped data for $Z > 80$ from Figs.~\ref{fig9} 
and \ref{fig10}, because the parameter $R$ is monotonous and very close to
1.

The E3 transitions affect radiative lifetime $\tau_{\mathrm{TOT}}$ of the level
$ 4p^64d^3\,( ^2H)4f\,\, ^3L_9$ not so significantly, as it can be seen from 
Fig.~\ref{fig10}. Nevertheless, their influence can not be neglected for  small 
$Z$ values. It should be mentioned that, for the zirconium isoelectronic 
sequence and for all other isoelectronic sequences investigated in current work,
the radiative lifetimes $\tau_{\mathrm{TOT}}$ for the levels with available M2 
transitions have larger $R$ values compared to the levels with only E3 
transitions available (see Fig.~\ref{fig5} and Fig.~\ref{fig6}).
Nevertheless, even if contribution of the E3 transitions to the calculated 
radiative lifetimes $\tau_{\mathrm{TOT}}$ is not so substantial, these 
transitions must be included into consideration, especially for the levels with 
the highest $J$ values when the M2 transitions are forbidden.

\subsection{Mo-like ions}
\label{mo}

\begin{figure}[!ht]
\includegraphics[scale=0.52]{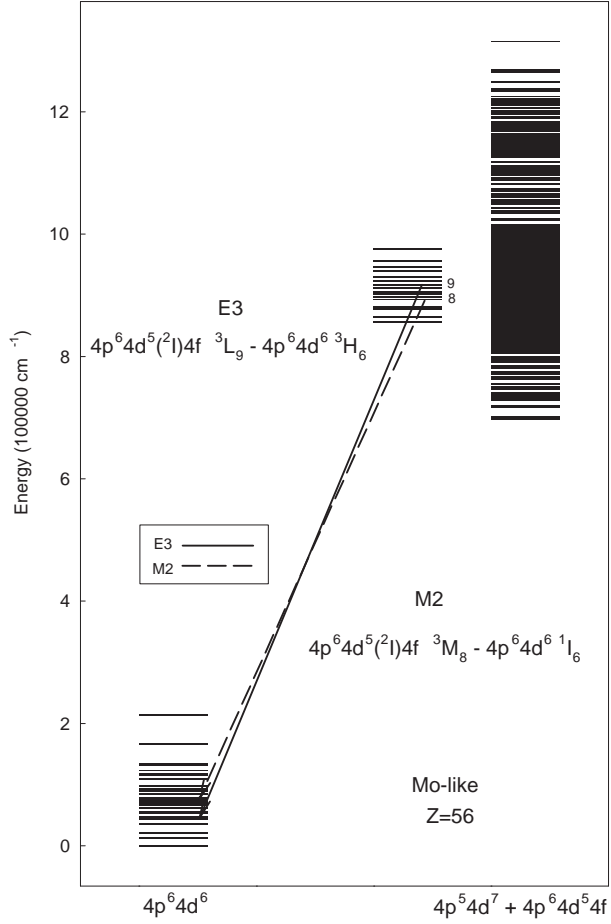}
\caption{
     \label{fig3} 
					Energy levels, E3 and M2 transitions for molybdenum-like barium ion
}
\end{figure}

Figure~\ref{fig3} demonstrates energy levels of investigated configurations 
for the molybdenum-like ion with $Z = 56$. The excited configuration levels 
with high total momentum $J$ values are presented in a separate column, 
to the left from the column which gives all energy levels of the odd-parity 
configurations (like in Fig.~\ref{fig1}).  There are no E1 transitions from 
these levels.  

The excited configurations $4p^54d^7$ and $4p^64d^54f$ of Mo-like ions have 
even larger number of metastable levels which have no decay channels through 
the E1 transitions. There are twelve levels with $J = 8$ and five levels with 
$J = 9$, see Fig.~\ref{fig3}. Moreover, there exists one level with $J=10$. 
Therefore, not only the E1 transitions are forbidden from this level.  
The M2 and E3 transitions also are forbidden, consequently, this metastable
level is not related to our study in general. However, it constitutes quite a 
special case. The lifetime $\tau_{\mathrm{TOT}}$ of this 
$4d^5\,(2I)\,4f\,\,^3M_{10}$ level is approximately equal to $0.05$\,s for 
$Z = 56$ and it decreases to $2.5 \cdot 10^{-6}$\,s for $Z=92$. This radiative 
lifetime $\tau_{\mathrm{TOT}}$ is defined by decay via two magnetic dipole 
transitions. The $J$ values for the ground configuration $4p^64d^6$ of this 
isoelectronic sequence is in range from 0 to 6.

\begin{figure}[!ht]
\includegraphics[scale=0.52]{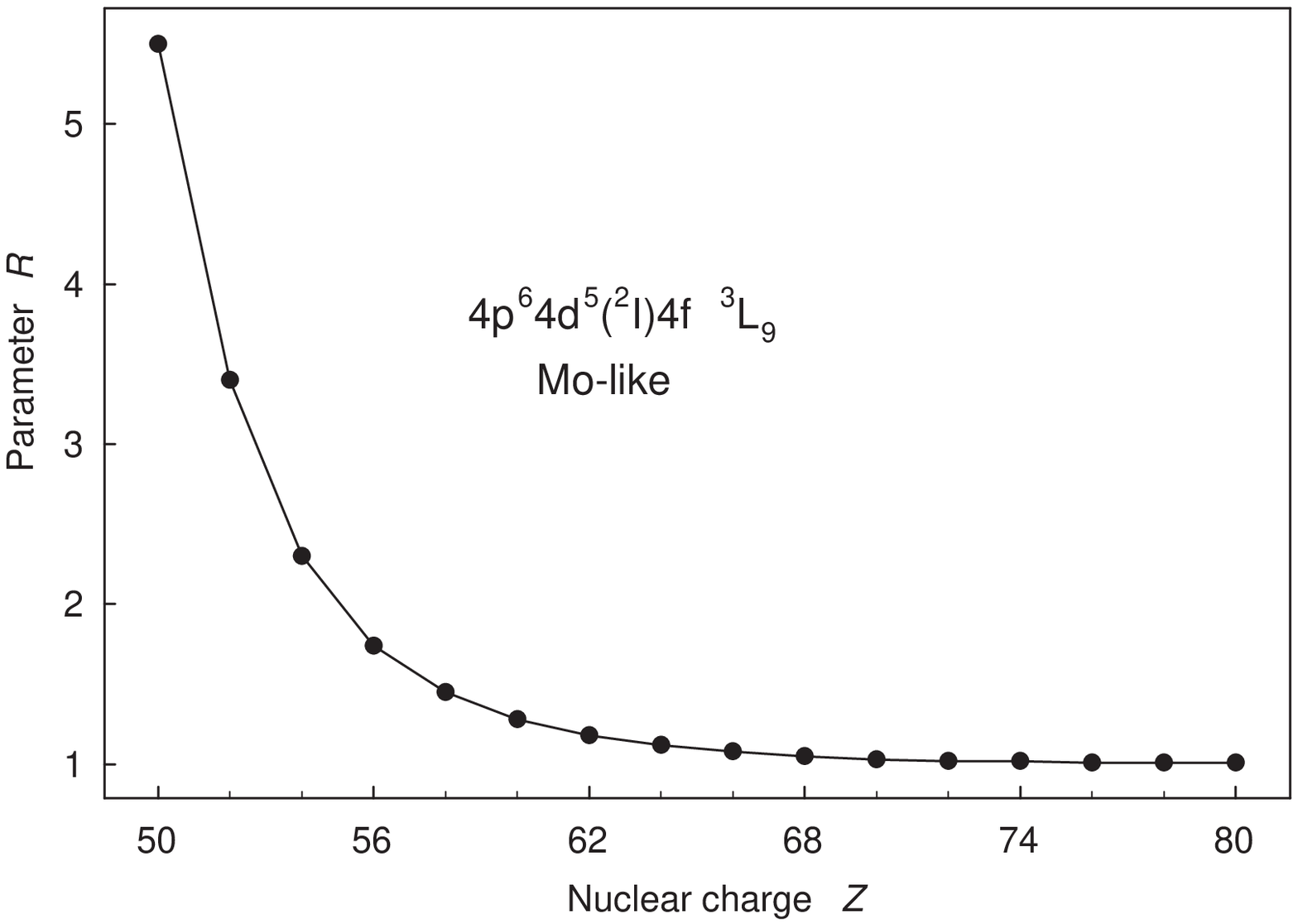}
\caption{
     \label{fig11} 
					Dependence of parameter $R$  on nuclear charge $Z$ for the level 
					$4p^54d^5\,( ^2I)4f\,\, ^3L_9$ of molybdenum-like ions
}
\end{figure} 

\begin{figure}[!ht]
\includegraphics[scale=0.52]{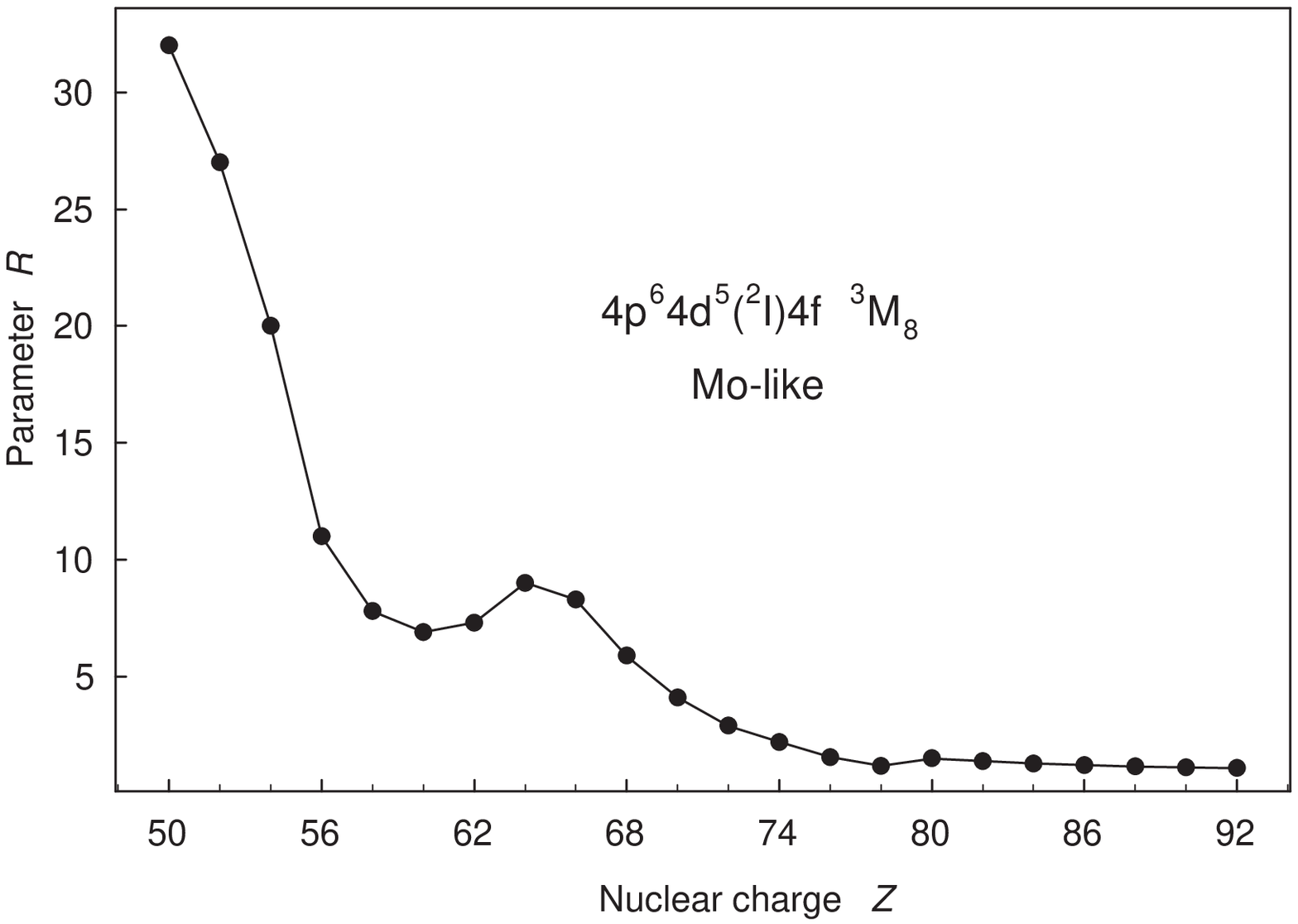}
\caption{
     \label{fig12} 
					Dependence of parameter $R$  on nuclear charge $Z$ for the level 
					$4p^54d^5\,( ^2I)4f\,\, ^3M_8$ of molybdenum-like ions
}
\end{figure} 

\begin{figure}[!ht]
\includegraphics[scale=0.52]{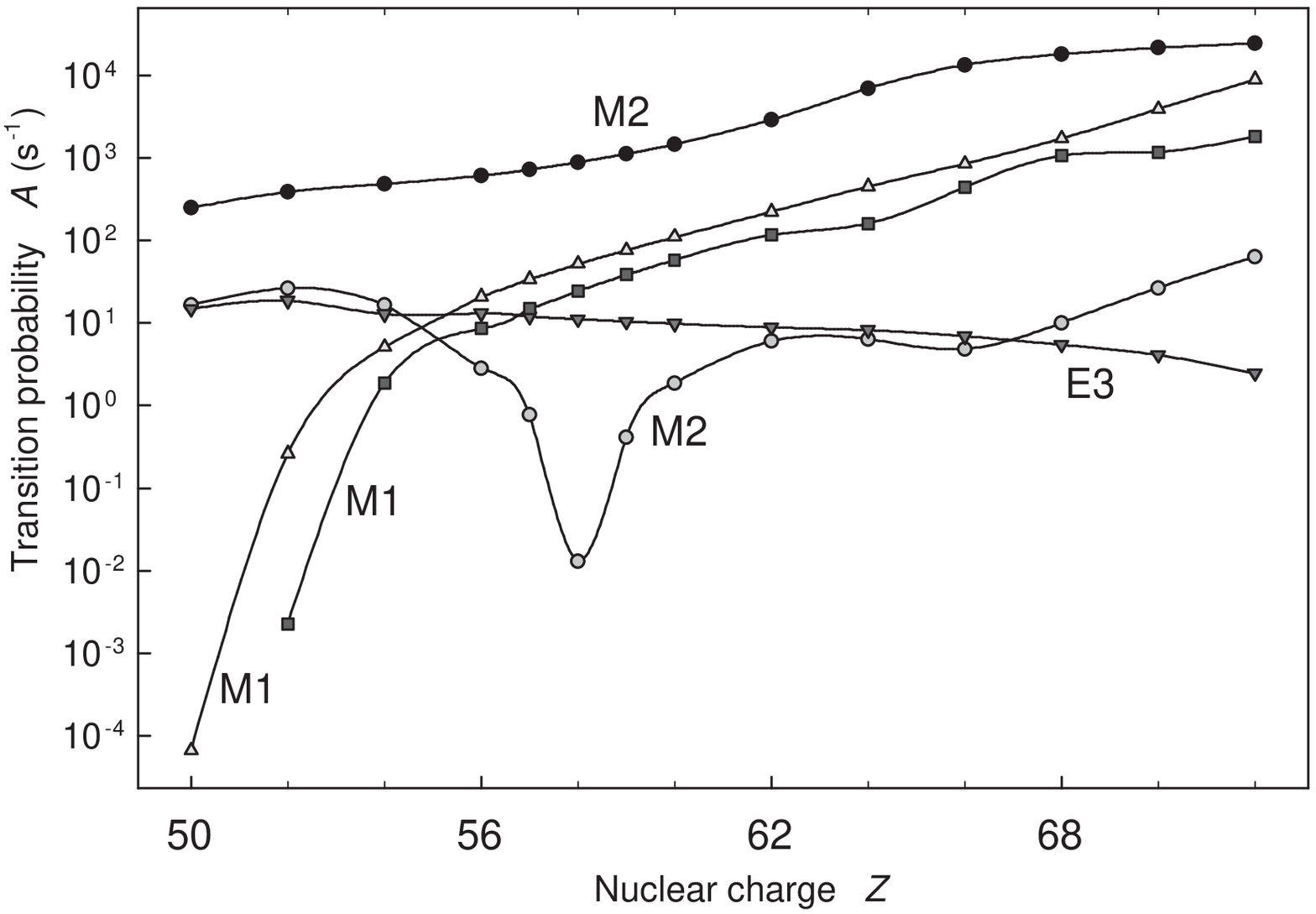}
\caption{
     \label{fig13} 
					Radiative transition probabilities originating from the 
					level $4p^64d^5\,( ^2I) 4f\,\,^3K_8$ along the molybdenum isoelectronic
					sequence. Two M1 transitions are presented by black squares and white
					triangles, two M2 transitions are presented by black and grey circles,
					E3 transitions are presented by grey triangles.					
}
\end{figure}

In the case of molybdenum isoelectronic sequence, we present the parameter $R$ 
for two levels. Figure~\ref{fig11} shows the dependence of parameter $R$ on 
nuclear charge $Z$ for the level $4p^64d^5\,( ^2I) 4f\,\, ^3L_9$. Beside the M1 
transitions, only E3 transitions are allowed from the level 
$4p^64d^5\,( ^2I) 4f\,\, ^3L_9$ to the ground configuration. Although the 
investigated configuration $4p^64d^54f$ has five levels with the total angular 
momentum $J = 9$, the mixing of terms is not significant here. Consequently, 
the parameter $R$ decreases very consistently for this configuration. 

A completely different behavior of $R$ for the level 
$4p^64d^5\,( ^2I) 4f\,\,^3K_8$ is presented in Fig.~\ref{fig12}. 
Radiative transition probabilities $A$ for this level, presented in 
Fig.~\ref{fig13}, clarify such an unusual behavior of the parameter $R$  along 
the isoelectronic sequence. Figure~\ref{fig13} presents only a part of 
isoelectronic sequence, ranging from $Z=50$ up to $Z=72$ because the transition 
probabilities dependence is smooth for higher $Z$ values. Furthermore, only the
strong transitions out of all possible ones from this level are given here. 
It is evident from this figure, that one M2 transition is stronger than all 
other, but probability of another M2 transition has a sharp minimum near $Z=58$. 
Whereas the probabilities of the M1 transitions evenly increase, this minimum of
the latter M2 transition causes the dip in Fig.~\ref{fig12}. 

A radiative transition probability $A$ is proportional to a square of the 
transition operator matrix element. When this matrix element changes its sign, 
its value becomes very close to a zero at some particular value of $Z$. 
Consequently, the square of the radiative transition operator matrix 
element has a sharp minimum at this $Z$, causing such a peculiar shape of 
$A(Z)$. Very similar behavior of radiative transition parameters proportional 
to a square of matrix element one can see in \cite{safr06}.

\subsection{Rh-like ions} 
\label{rh}

\begin{figure}[!ht]
\includegraphics[scale=0.52]{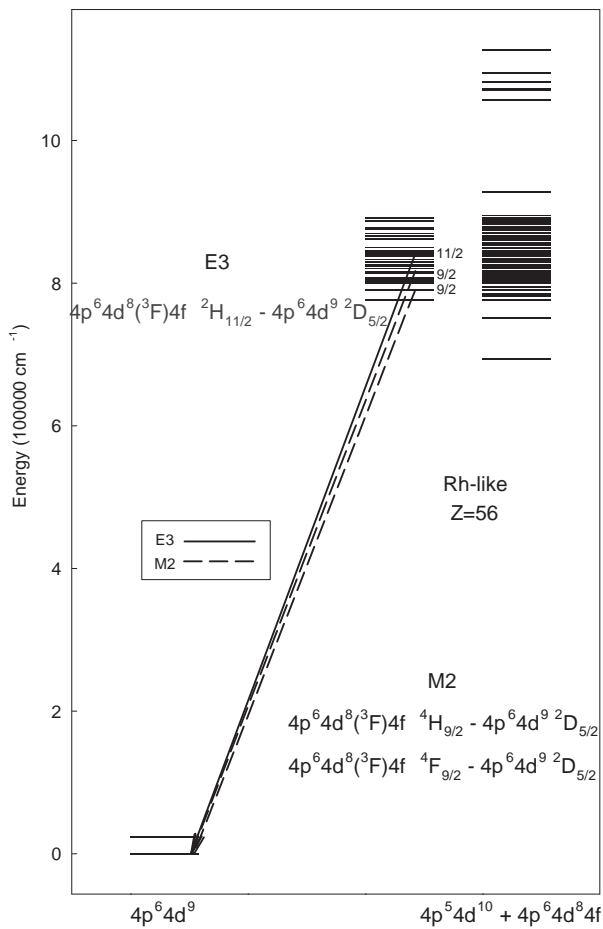}
\caption{
     \label{fig4} 
					Energy levels, E3 and M2 transitions for rhodium-like barium ion 
}
\end{figure}

Figure~\ref{fig4} displays the energy level structure of investigated 
configurations  for the rhodium-like ion with $Z = 56$. These ions are the most 
complex ones, with regard to metastable levels, when considering sequences with
open $4d$ shell. There is a large number of excited levels having large $J$ 
values, whereas the ground configuration is $4p^64d^9\,\, ^2D$. So, only two 
levels, $J=3/2$ and $J=5/2$ make up the ground configuration. 
We have placed into a separate column all the levels with $J \geq 9/2$ from the
configuration complex $4p^54d^{10} + 4p^64d^84f$. There are two levels with 
$J=15/2$, five levels with $J=13/2$, nine $J=11/2$ levels and thirteen levels 
with $J=9/2$. These 29 levels make up one third of complete set of excited 
odd-configuration levels. Not only E1, but also E3 and M2 transitions are 
forbidden from the levels with $J=13/2$ and $J=15/2$. Therefore these levels 
can decay only through transitions inside their configuration complex. 
The properties of these levels will be  discussed further in this section.
The significance of the transitions from three levels presented in 
Fig.~\ref{fig4} is studied further in this section where the radiative 
lifetimes $\tau_{\mathrm{TOT}}$ of these metastable and some other levels and 
their  possible decay channels are discussed too.

\begin{figure}[!ht]
\includegraphics[scale=0.52]{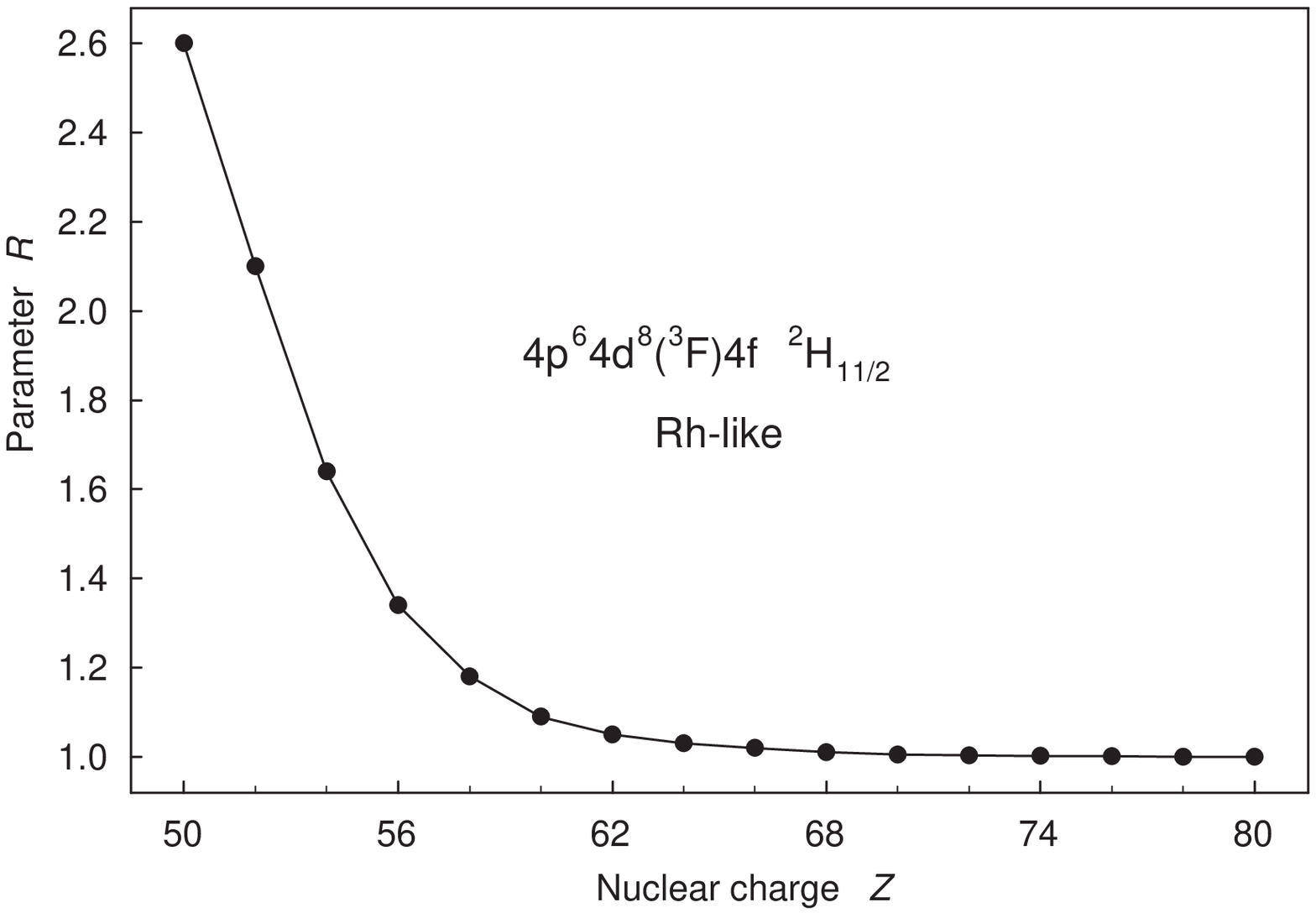}
\caption{
     \label{fig14} 
					Dependence of parameter $R$  on nuclear charge $Z$ for the level 
					$4p^54d^8\,( ^3F)4f\,\, ^2H_{11/2}$ of rhodium-like ions
}
\end{figure} 

\begin{figure}[!ht]
\includegraphics[scale=0.52]{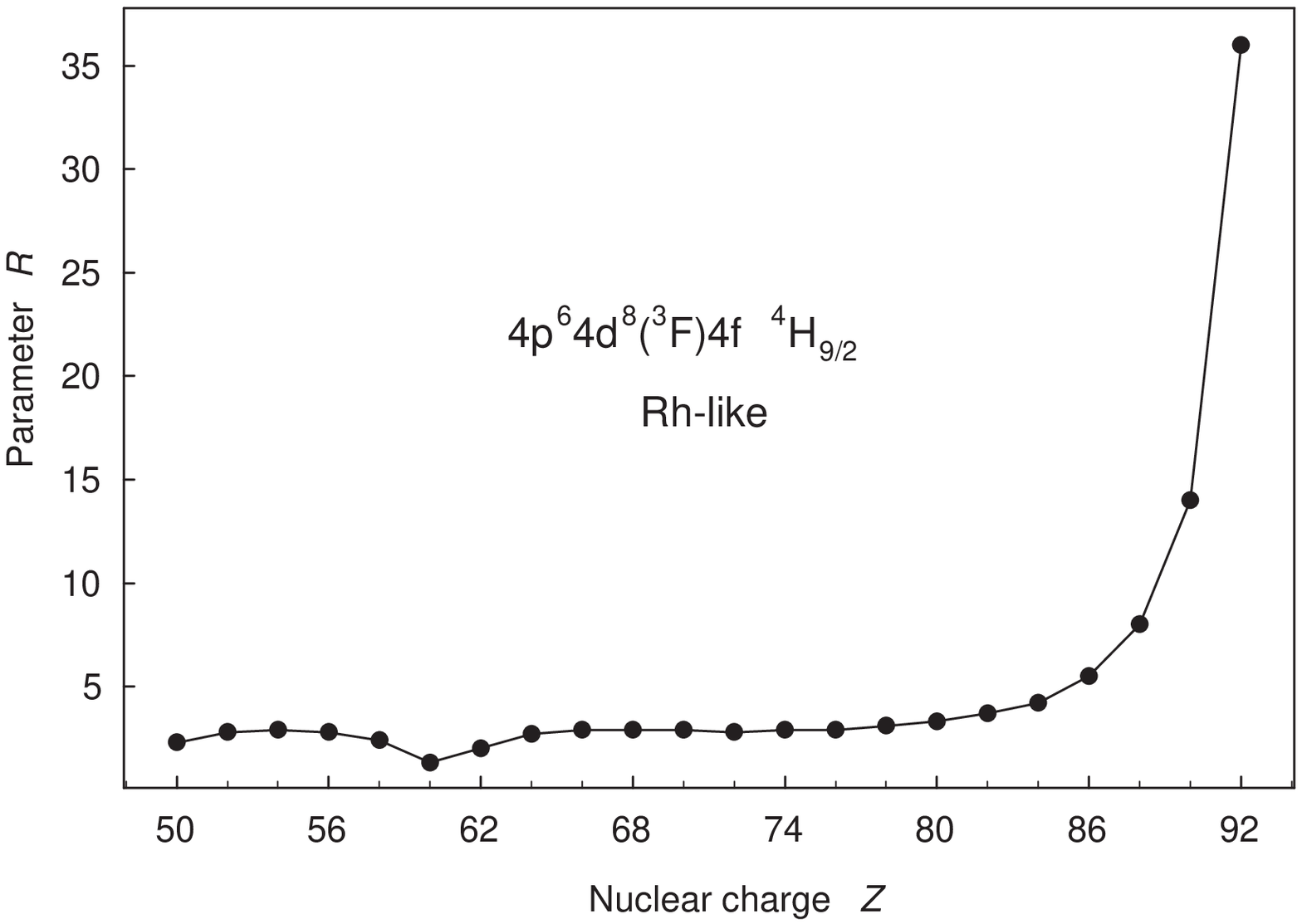}
\caption{
     \label{fig15} 
					Dependence of parameter $R$  on nuclear charge $Z$ for the level 
					$4p^54d^8\,( ^3F)4f\,\, ^4H_{9/2}$ of rhodium-like ions
}
\end{figure} 

\begin{figure}[!ht]
\includegraphics[scale=0.52]{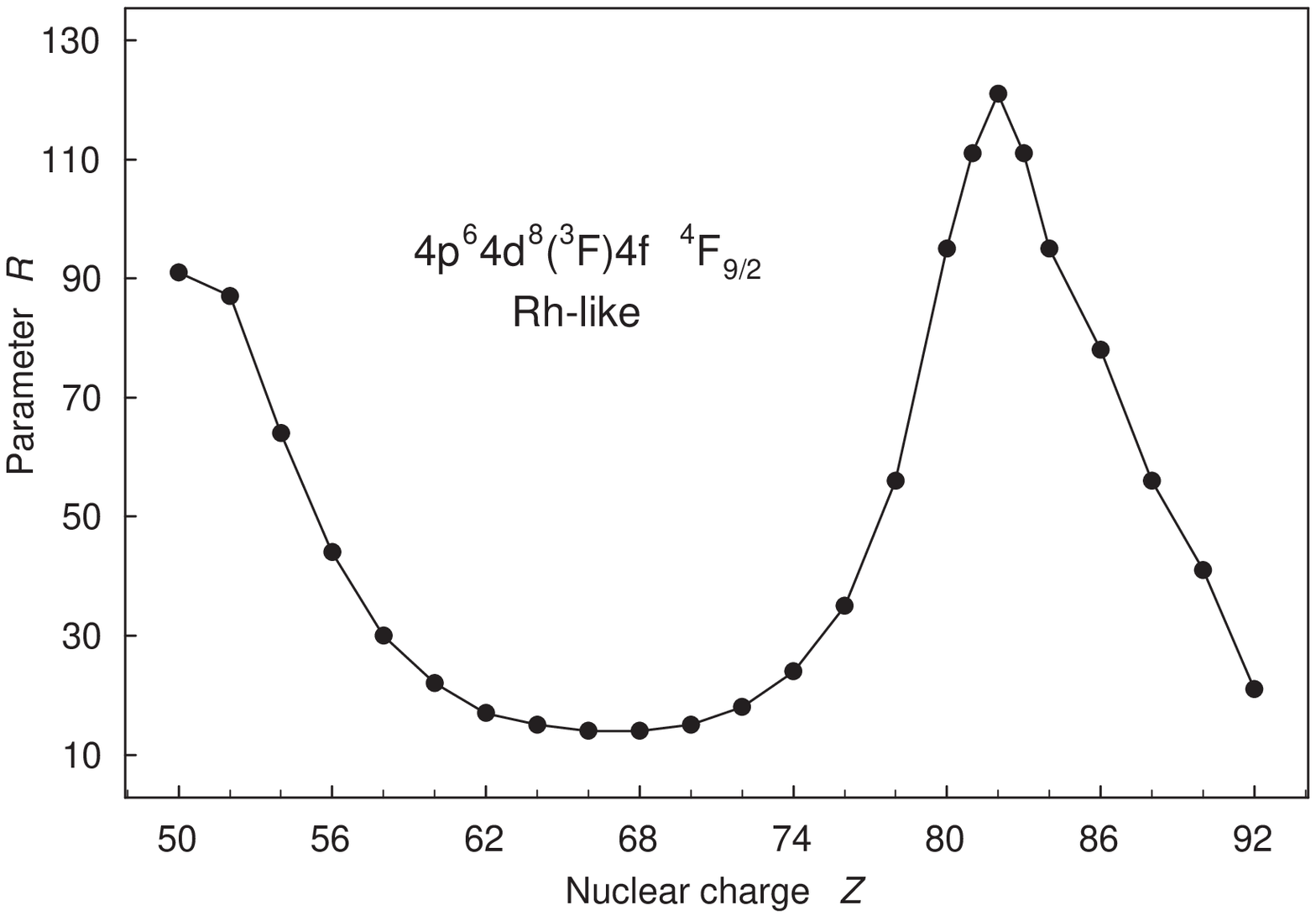}
\caption{
     \label{fig16} 
					Dependence of parameter $R$  on nuclear charge $Z$ for the level 
					$4p^54d^8\,( ^3F)4f\,\, ^4F_{9/2}$ of rhodium-like ions
}
\end{figure} 

\begin{figure}[!ht]
\includegraphics[scale=0.52]{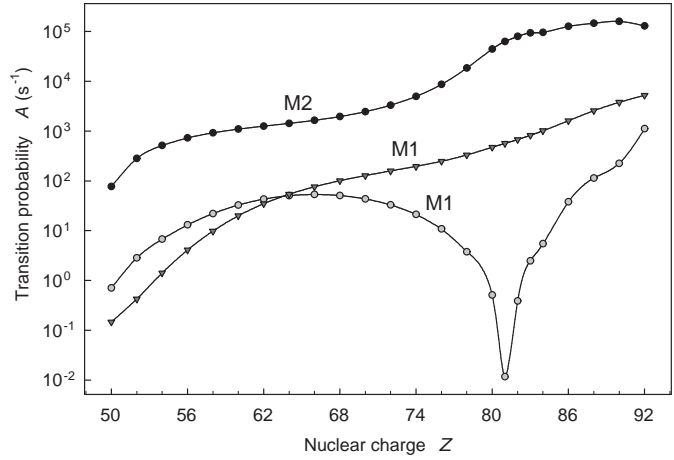}
\caption{
     \label{fig17} 
					Radiative transition probabilities originating from the 
					level $4p^54d^8\,( ^3F)4f\,\, ^4F_{9/2}$ along the rhodium isoelectronic
					sequence. Two M1 transitions are presented by grey circles and grey 
					triangles, M2 transitions are presented by black circles
}
\end{figure} 

Properties of the energy levels for the ions from the rhodium isoelectronic 
sequence are presented in Figs.~\ref{fig14}, \ref{fig15}, \ref{fig16}. 
Only the E3 transitions to the ground configuration are allowed from the level 
$4p^64d^8\,( ^3F)4f\,\,^2H_{11/2}$ (Fig.~\ref{fig14}). The value of $R$
is comparatively small and it decreases rapidly. For the  
$4p^64d^8\,( ^3F)4f\,\,^4H_{9/2}$ level, the parameter $R$ increases sharply 
(see Fig.~\ref{fig15}) only at the end of the sequence. Quite a large number 
of transitions contribute to the defined radiative lifetime 
$\tau_{\mathrm{TOT}}$ value of this level, such as strong M2 transition, two 
weak E3 transitions and several different M1 transitions. A sharp increase of 
$R$ is caused by the M2 transition probability being larger than that of the M1 
transitions and increasing significantly faster. A small irregularity at $Z=60$ 
appears due to a minimum of the dominant  M2 transition probability value at 
these $Z$.

The most interesting behavior of the parameter $R$ can be seen in 
Fig.~\ref{fig16}, presenting the level $4p^54d^8\,( ^3F)4f\,\, ^4F_{9/2}$. 
To explain it, we present calculated radiative transition probabilities from 
this level in Fig.~\ref{fig17}. In the first part of the isoelectronic sequence, 
$R$ demonstrates quite an usual behavior. Here the influence of the strongest 
M2 transition decreases, although the probability values of that transition 
increase. Nevertheless, this increase is slower comparing to increase of the 
M1 transition probability values. When $Z$ is close to 70, the M1 transition 
probability values begin to decrease and they reach minimum at $Z=81$. The 
decrease and subsequent increase is so sharp that we have performed calculations
for several additional ions in the proximity of these $Z$ values. As it is 
mentioned in previous sections, such a dependence of the radiative transition 
probabilities is caused by the behavior of the corresponding radiative 
transition operator matrix element. For other levels with $J=9/2$ and $J=11/2$, 
the parameter $R$ values are small, because these levels are located 
comparatively high in energy spectra and have strong decay channels through M1 
transitions.

As it was mentioned above, only the transitions inside excited configurations
complex are allowed from levels  with $J=13/2$ and $J=15/2$. In particular 
parts of the rhodium sequence, even M1 and E2 transitions are not
possible for these metastable levels.  This is caused by the 
location of these levels in energy spectra. The level with the highest $J$ 
value of $4p^64d^8\,( ^3F)4f\,\,^4I_{15/2}$ is too low to have any 
radiative transition when $50 \leq Z \leq 60$, because the lower energy levels 
only have $J < 9/2$. For the higher-$Z$ ions, the 
$4p^64d^8\,( ^3F)4f\,\, ^2I_{13/2}$ level energy is lower than that of
$^4I_{15/2}$, therefore rather weak M1 transition becomes possible. 
The determined radiative lifetime of this level $\tau = 2500$~s
for $Z=52$  and decreases to $\tau = 5.7 \cdot 10^4$~ns for $Z=92$.
 
The level $4p^64d^8\,( ^3F)4f\,\,^2I_{13/2}$ does not have a radiative
decay channel when $82 \leq Z \leq 92$. When $Z=80$, only one level with $J$ 
value high enough, namely, $4p^64d^8\,(3F)4f\,\,^2G_{9/2}$ lies below the level
with $J=13/2$. Therefore, the E2 radiative transition becomes possible, but the
calculated radiative lifetime $\tau_{\mathrm{E2}} = 1.5 \cdot 10^9$~s. This 
equals to approximately 46 days. The low lying level 
$4p^64d^8\,( ^3F)4f\,\,^2G_{9/2}$ also has some peculiar properties. For 
$Z < 66$, it has only one radiative decay channel via the M2 transition. When 
$Z \geq 66$, a decay channel through the M1 radiative transition is open, but 
due to a negligible transition probability values, it becomes somehow 
significant only when $Z = 86$. That means that only the magnetic quadrupole 
transition is meaningful for this level along all isoelectronic sequence. 

There is one more quite unusual level $4d^8\,(^3F)\,4f\,\,^2S_{1/2}$ in this
isoelectronic sequence. Formally, it does not satisfy our criteria, because
the E1 transitions are allowed from it. Following comprehensive studies, we 
have noticed that, beside electric dipole transition to the $^2D_{3/2}$
level of the ground configuration, the M2 transition to the $^2D_{5/2}$
level becomes significant starting from $Z=78$. When $Z$ increases, 
the value of M2 transition probability becomes similar to that of E1 transition.

\subsection{Remaining $4d^N$ sequences} 
\label{other}

We must highlight that in current work we present only a small part of our
results. We have investigated metastable levels in all isoelectronic sequences
with $4d^N$ shell in the ground configuration. Magnetic dipole transitions are
the most common decay channel for the levels of excited configurations.
In previous sections we have described the cases when the E3 and M2 radiative 
transitions become significant decay channels.  As a rule, when the M2 
radiative transitions are possible, these are much stronger comparing to the 
E3 transitions originating from the same level. In some particular cases, 
the E3 transition probabilities can be of the same magnitude as those of the 
M2 transitions. Moreover, there are energy levels with the highest $J$ values,
which can not decay by the M2 transitions.

Furthermore, the radiative decay is possible to other lower levels with a 
suitable $J$ value, not only to those given in our plots. It was established, 
that, for some levels, the M2 and E3 transitions are significant only in a 
narrow $Z$ range of their isoelectronic sequence. In current work, we present 
the most typical dependences, where the parameter $R$ decreases when nuclear 
number $Z$ increases. Other particular cases, when $R$ increases with $Z$ 
increase, are given here too along with some more peculiar behavior of the 
parameter $R$. All these statements are valid both for the sequences described 
in Sects.~\ref{sr}, \ref{zr}, \ref{mo}, \ref{rh} and for remaining $4d^N$ 
sequences with $N=1,3,5,7,8,10$.

The ground configuration for Rb-like ions is $4p^64d\,\, ^2D$. The first two
excited configurations do not have energy levels with significant M2 and E3
transitions. This is caused by the fact that levels with maximum $J$ values are 
located comparatively high, therefore M1 and E2 transitions to the lower levels 
of the same configurations are allowed. Only the level 
$4p^54d^2\,(^3F)\,\,^4F_{9/2}$ has $R > 2$ at the beginning of the sequence, but 
it becomes $R < 1.1$, when $Z = 64$.

The ground configuration for yttrium-like ions is $4p^64d^3$. The M2 and E3 
transitions do not play an important role here because the configuration 
$4p^64d^3$ has levels with large $J$ values ($J_{\mathrm{max}}=11/2$). Only 
four excited levels have $J=15/2$ and just two of them have the M2 and E3 
transitions with significant probability values. For the level
 $4d^2\,(^3F)4f\,\, ^4I_{15/2}$, the $R > 12$ at the beginning of the 
isoelectronic sequence when $Z = 52$ but it decreases sharply when $Z$ 
increases. Such a behavior is caused by one M2 transition with transition 
probability value more than 100 times larger than the value of strongest M1 
transition  at the beginning of isoelectronic sequence. At the top end of the
isoelectronic sequence, the M2 transition probability becomes similar to that 
of several M1 transitions which determine radiative lifetime of this level.
Another level, namely $4p^54d^4\,(^3H)\,\, ^4I_{15/2}$, has 
$R_{\mathrm{max}} = 4$ at the beginning of the isoelectronic sequence. 
For the higher-$Z$ ions, $R$ decreases rather slowly comparing to other levels.
Such a behavior is caused by fact that decay of this level is going through many
channels. There are three E3 transitions and one M2 transition to the ground 
configuration levels in addition to M1 and E2 transitions among the levels of 
this configuration. Moreover, this level is rather special one, because at 
higher $Z$ values, the E2 transitions become more important than M1 transitions 
for the calculated radiative lifetime $\tau_{\mathrm{TOT}}$, contrary to the most 
levels in this and other investigated isoelectronic sequences.

The ground configuration for niobium isoelectronic sequence is $4p^64d^5$. 
The excited configuration levels with $J=19/2$ are located rather high in 
energy level spectra. Therefore, strong M1-type transitions are possible to 
lower-lying levels with $J=17/2$. There are six levels with $J=17/2$, but only 
two of them have their radiative lifetimes significantly affected  by 
transitions of higher multipolar order. For the first of them,  
$4d^4\,(^3H)4f\,\, ^4K_{17/2}$, the parameter $R$ gradually decreases from 
the top value of $R = 255$ with $Z$ increase. The decay of this level is going 
mainly through a strong M2 transition. Meanwhile, the parameter $R$ changes 
non-monotonously along the $Z$ and has a maximum value of $R=39$ at $Z=72$ for 
the level $4d^4\,(^3H)4f\,\, ^2L_{17/2}$. Such a non-standard behavior can be 
explained in a similar way as it has been done for Mo-like ions 
(Fig.~\ref{fig12}) and for Rh-like ions (Fig.~\ref{fig16}).

The ground configuration for technetium-like ions is $4p^64d^7$. For the
excited-configuration levels with $J = 17/2$, only the E3 transitions are 
allowed. Furthermore, these transitions are weak and have some significance 
only for the level $4d^6\,(^3H)4f\,\, ^2L_{17/2}$ at the lower end of this 
isoelectronic sequence. The parameter $R = 4$ at $Z = 52$, and it decreases 
sharply for higher $Z$ values. There are fifteen energy levels with $J=15/2$, 
but only two of them are significantly affected by the M2 and E3 radiative 
transitions. For other levels, the influence of these transitions is 
insignificant even at the low end of the isoelectronic sequence. 

In the case of $4d^6\,(^3H)4f\,\, ^2K_{15/2}$ level, inclusion of the M2 decay 
channel changes the value of radiative lifetime $\tau_{\mathrm{TOT}}$ more than 
4000 times at the beginning of the isoelectronic sequence. It decreases fast down
to $R=7$ at $Z=76$. On the other hand, the M2 transition is not so important for
the level $4d^6\,(^3H)4f\,\, ^4I_{15/2}$. The parameter $R$ is approximately 
equal to 40 at the beginning of the sequence. It decreases rather slowly, 
and $R=2$ when $Z=76$.

The ground configuration for Ru-like ions is $4p^64d^8$. The total angular 
momentum $J$ maximum values can be 9 and 8 for the excited configurations 
$4p^64d^74f$ and $4p^54d^8$. The E3 and M2 transitions from these levels are 
forbidden. They have one or more M1 transitions to lower levels  with $J=8$ and
$J=7$ from the same configuration complex. 

An interesting case constitutes the level $4d^7\,(^4F)4f\,\, ^3I_7$. The main 
decay channels for this level are the radiative E3 and M1 transitions. The 
parameter $R$ increases from the beginning of the isoelectronic sequence (like 
in Fig.~\ref{fig7}) and reaches maximum value of $R=30$ at $Z=92$. 

There are nineteen $J=6$ levels, most of them having radiative lifetimes 
$\tau_{\mathrm{TOT}}$ significantly affected by transitions of higher multipole 
orders. This is especially noticeable at the beginning of isoelectronic sequence
where the parameter $R$ values are close to 10. Very high $R$ values were
determined for three lowest energy levels with $J = 6$. The level 
$4d^7\,(^4F)4f\,\, ^5G_6$ is very special one because of possible M2 transition 
being virtually the only available decay channel (except for very weak E2 
transition), therefore $R > 900,000$ when $Z = 52$. The value of $R$ becomes 
smaller than 1000 only starting with $Z>80$ ions. Only in the case, when some 
level with $J=5$ has an energy lower than this $4d^7\,(^4F)4f\,\, ^5G_6$ level, 
the decay through the M1 transition becomes possible. For the levels 
$4d^7\,(^4F)4f\,\, ^3H_6$ and $4d^7\,(^2G)4f\,\, ^1I_6$, the $R$ values reach 
30 and 40, correspondingly, at the beginning of this isoelectronic sequence, but 
they decrease sharply for higher-$Z$ ions.

The ground configuration of palladium isoelectronic sequence is $4p^64d^{10}$.
Therefore, only one excited configuration $4p^64d^94f$ can be considered. Only 
three levels from this excited configuration can decay through E1 transitions, 
because the ground configuration has the only $^1S_0$ level. Two levels, 
$4d^9\,(^2D)4f\,\, ^3P_0$ and $4d^9\,(^2D)4f\,\, ^3H_0$, have no radiative decay
channels at all. In general, the radiative lifetimes of this configuration 
levels are determined by M1 transitions. Certainly, E2 transitions are allowed, 
too, but their transition probabilities are significantly smaller than those of 
M1 transitions. However, there are levels where transitions of higher multipole
order play an important role. Four levels with $J=2$ can decay not only through 
M1 transitions but also via M2 transitions. This channel is extremely important
for the levels $4d^9\,(^2D)4f\,\, ^3F_2$ and $4d^9\,(^2D)4f\,\, ^3P_2$. 
The parameter $R$ decreases from 800 to 10 when $Z$ changes from 52 to 92,
i.e. over all range of examined ions, for the $^3F_2$ level and from 1400 to 5 
for the $^3P_2$ level. For other two levels with $J=2$, the influence of M2 
transitions on total radiative lifetime $\tau_{\mathrm{TOT}}$ is not so very 
large but still rather noticeable. The M1 and E3 transitions are allowed from 
the levels with $J=3$. The E3 transitions are not significant comparing to M2 
transitions. Therefore $R$  values are close to 1. Only the level
$4d^9\,(^2D)4f\,\, ^1F_3$ makes some exception having $R$ close to 4 at $Z=52$.
Nevertheless, it decreases fast and reaches 1.4 at $Z=60$.

\section{Conclusions}

The magnetic quadrupole and electric octupole transitions 
from some levels of the first excited configurations $4p^54d^{N+1}$ and 
${\mathrm 4p^64d^{N-1}4f}$ to the ground configuration $4p^64d^N$ can 
play very significant role for their radiative lifetimes. These levels have 
large total angular momentum $J$ values, therefore the radiative decay to the 
ground configuration via electric dipole transitions is not allowed. 
Consequently, these levels are metastable ones, therefore one assumes that
the M1 and E2 transitions shall become the main decay channels and determine 
the lifetimes of such excited levels. Nevertheless, since such levels are 
located in the lower part of their own configuration complex energy spectra,
possible transitions (M1 and E2) among the excited configuration 
$4p^54d^{N+1}$ and ${\mathrm 4p^64d^{N-1}4f}$ levels are severally restricted.
Our investigation has demonstrated that radiative transitions of higher 
multipole order, such as M2 and E3, become significant.

Inclusion of additional decay through radiative M2 and E3 transition channels
decreases values of determined radiative lifetimes $\tau_{\mathrm{TOT}}$.
The main impact on calculated radiative lifetimes $\tau_{\mathrm{TOT}}$ 
originates from M2 transitions if these are allowed by $J$ selection rules, 
whereas E3 transition probability values are smaller.  
The E3 transitions become significant when $J$ acquires maximum values and 
therefore the M2 transitions are forbidden.

The influence of E3 and M2 transitions can both to decrease and to increase 
along the isoelectronic sequence. This originates both from the variation of 
transition probability values and from the location changes of investigated 
levels in respect to other levels of the same configuration and variation 
of corresponding M1 and E2 transition probabilities.

Most of investigated levels belong to configurations $4p^64d^{N-1}4f$. 
It is simple to explain that such an excitation always allows to get large 
values of the final $LS$ momenta and the total momentum $J$. For the excitation 
of the $4p$ electrons to the states with $4d$ electrons, large total angular
momenta $J$ appear only in the case when a number of electrons in $4d$ shell 
is small. When $N$ is large ($N=9,10$), there exists levels with 
forbidden both E1, E2, E3 and M1,M2 transitions.

Performed investigation has demonstrated that it is difficult to predict the
significance of the E3 and M2 transitions without performing detailed 
calculations. This is consequence of the fact that the values of different 
radiative transition probabilities can change dramatically along the 
isoelectronic sequences due to changes of the eigenfunctions and transition 
energies.

\begin{acknowledgments}

Current research is funded by the European Social Fund under the Global Grant 
measure, project VP1-3.1-{\v S}MM-07-K-02-013. 

\end{acknowledgments}


\providecommand{\newblock}{}


\begin{thebibliography}{99}
\expandafter\ifx\csname natexlab\endcsname\relax\def\natexlab#1{#1}\fi
\expandafter\ifx\csname bibnamefont\endcsname\relax
  \def\bibnamefont#1{#1}\fi
\expandafter\ifx\csname bibfnamefont\endcsname\relax
  \def\bibfnamefont#1{#1}\fi
\expandafter\ifx\csname citenamefont\endcsname\relax
  \def\citenamefont#1{#1}\fi
\expandafter\ifx\csname url\endcsname\relax
  \def\url#1{\texttt{#1}}\fi
\expandafter\ifx\csname urlprefix\endcsname\relax\def\urlprefix{URL }\fi
\providecommand{\bibinfo}[2]{#2}
\providecommand{\eprint}[2][]{\url{#2}}

\bibitem{pb12} 
P. Bogdanovich and R. Kisielius, 
At. Data Nucl. Data Tables, (2013) http://dx.doi.org/10.1016/j.adt.2012.11.001 (in press).

\bibitem{pb13} 
P. Bogdanovich and R. Kisielius, 
At. Data Nucl. Data Tables, (2012) (submitted).

\bibitem{biemont2004}
E. Bi\'emont, A. Derkatch, P. Lundin, S. Mannervik, L.-O. Norlin, D. Rostohar, 
P. Royen, P. Palmeri, and P. Schef,
Phys. Rev. Lett., {\bf 93}, 063003 (2004).

\bibitem{cff06} 
C. Froese Fischer, G. Tachiev, and A. Irimia, 
At. Data Nucl. Data Tables, {\bf 92}, 607 (2006).

\bibitem{safr06} 
U. I. Safronova, A. S. Safronova, S. M. Hamasha, and P. Beiersdorfer, 
At. Data Nucl. Data Tables, {\bf 92}, 47 (2006).

\bibitem{safr09} 
U. I. Safronova and M. S. Safronova, 
Phys. Rev. A {\bf 79}, 032511 (2009).

\bibitem{lun07} 
P. Lundin, J. Gurell, L.-O. Norlin, P. Royen, S. Mannervik, P. Palmeri, 
P. Quinet, V. Fivet, and E. Biemont, 
Phys. Rev. Lett. {\bf 99}, 213001 (2007).

\bibitem{lun08} 
P. Lundin, J. Gurell, S. Mannervik, P. Royen, L.-O. Norlin, H. Hartman, 
and A. Hibbert, 
Phys. Scr. {\bf 78}, 015301 (2008).

\bibitem{bie07} 
E. Biemont, A. Ellmann, P. Lundin, S. Mannervik, L.-O. Norlin, P. Palmeri, 
P. Quinet, D. Rostohar, P. Royen, and P. Schef, 
Eur. Phys. J. D {\bf 41}, 211 (2007).

\bibitem{pbor06} 
P. Bogdanovich and O. Rancova, 
Phys. Rev. A {\bf 74}, 052501 (2006).

\bibitem{pbor07} 
P. Bogdanovich and O. Rancova, 
Phys. Rev. A {\bf 76}, 012507 (2007).

\bibitem{pbor02} 
P. Bogdanovich and O. Rancova, 
Lithuan. J. Phys. {\bf 42}, 257 (2002).

\bibitem{pbor03} 
P. Bogdanovich and O. Rancova, 
Lithuan. J. Phys. {\bf 43}, 177 (2003).

\bibitem{pbvjor05} 
P. Bogdanovich, V. Jonauskas, and O. Rancova, 
Nucl. Instr. Meth. B {\bf 235}, 145 (2005).

\bibitem{hfr} 
R.D. Cowan, 
{\sl The Theory of Atomic Structure and Spectra}, 
(University of California Press, Los Angeles, 1981).

\bibitem{pbor08} 
P. Bogdanovich and O. Rancova, 
Phys. Scr. {\bf 78}, 045301 (2008).

\bibitem{cff91a} 
A. Hibbert, R. Glass, and C. F. Fischer, 
Comput. Phys. Commun. {\bf 64}, 445 (1991).

\bibitem{cff91b} 
C. F. Fischer, M. R. Godefroid, and A. Hibbert, 
Comput. Phys. Commun. {\bf 64}, 486 (1991).

\bibitem{cff91c} 
C. F. Fischer and M. R. Godefroid, 
Comput. Phys. Commun. {\bf 64}, 501 (1991).


\end{thebibliography}
\end{document}